\documentclass[aps,prd,preprint,tightenlines,groupedaddress,nofootinbib]{revtex4}
\usepackage{amssymb,latexsym}
\usepackage{amsmath,amsbsy}
\usepackage{epsfig,bm}
\usepackage{graphicx,comment}
\DeclareMathOperator{\tr}{tr}

\begin{document}
\def\a{{\alpha}}
\def\b{{\beta}}
\def\d{{\delta}}
\def\D{{\Delta}}
\def\e{{\epsilon}}
\def\g{{\gamma}}
\def\G{{\Gamma}}
\def\k{{\kappa}}
\def\l{{\lambda}}
\def\L{{\Lambda}}
\def\m{{\mu}}
\def\n{{\nu}}
\def\o{{\omega}}
\def\O{{\Omega}}
\def\S{{\Sigma}}
\def\s{{\sigma}}
\def\th{{\theta}}

\def\ol#1{{\overline{#1}}}

\def\Dslash{D\hskip-0.65em /}

\def\CPT{{$\chi$PT}}
\def\QCPT{{Q$\chi$PT}}
\def\PQCPT{{PQ$\chi$PT}}
\def\tr{\text{tr}}
\def\str{\text{str}}
\def\diag{\text{diag}}
\def\order{{\mathcal O}}

\def\cC{{\mathcal C}}
\def\cB{{\mathcal B}}
\def\cT{{\mathcal T}}
\def\cQ{{\mathcal Q}}
\def\cL{{\mathcal L}}
\def\cO{{\mathcal O}}
\def\cA{{\mathcal A}}
\def\cQ{{\mathcal Q}}
\def\cR{{\mathcal R}}

\def\eqref#1{{(\ref{#1})}}

\preprint{NT@UW 03-018}
\title{Electromagnetic Properties of the Baryon Decuplet in \\ Quenched and Partially
Quenched Chiral Perturbation Theory}
\author{Daniel Arndt}
\email[]{arndt@phys.washington.edu}
\author{Brian C.~Tiburzi}
\email[]{bctiburz@phys.washington.edu}
\affiliation{Department of Physics,  	Box 351560,\\
	University of Washington,     
	Seattle, WA 98195-1560, USA}
\date{\today}

\begin{abstract}
We calculate electromagnetic properties of the decuplet baryons in quenched and partially 
quenched chiral perturbation theory. We work at next-to-leading order in the chiral 
expansion, leading order in the heavy baryon expansion, and obtain expressions
for the magnetic moments, charge radii, and electric quadrupole moments. 
The quenched calculation is shown to be pathological since only quenched chiral singularities 
are present at this order. We present the partially quenched results for 
both the $SU(2)$ and $SU(3)$ flavor groups and use the isospin limit in the latter. These 
results are necessary for 
proper extrapolation of lattice calculations of decuplet electromagnetic properties. 
\end{abstract}

\maketitle

\section{Introduction}

The study of hadronic electromagnetic properties
provides important insight into the non-perturbative
structure of QCD.
A model-independent tool to study QCD at low energies is
chiral perturbation theory (\CPT), 
which is an effective field theory
with low-energy degrees of freedom, e.g., the meson octet
in $SU(3)$ flavor.
\CPT\ assumes that these mesons are the pseudo-Goldstone
bosons that appear from spontaneous breaking of chiral symmetry
from
$SU(3)_L\otimes SU(3)_R$ down to
$SU(3)_V$. 
Observables receive contributions from both
long-range and short range physics;
in \CPT\ the long-range contribution arises from
the (non-analytic) pion loops, while the short-range
contribution arises from low-energy constants. The number of these low-energy constants is constrained by 
symmetries but their values must be determined from experiment or lattice simulations.

Progress in measuring the 
proton and neutron form factors
has been made (see \cite{Mergell:1996bf,Hammer:1996kx} for references), 
including recent high precision measurements for the proton~%
\cite{Gayou:2001qd}.
Experimental study of the remaining baryons, in particular the decuplet 
baryon resonances, however, is much harder. Experiments measuring the decuplet 
magnetic moments are anticipated in the foreseeable future. The Particle Data Group
lists values for the $\D^{++}$ magnetic moment~%
\cite{Hagiwara:2002fs} but with 
sizeable discrepancy and uncertainty, even among the two most recent results~%
\cite{Bosshard:1991zp,LopezCastro:2000ep}. A report of the initial measurement of 
the $\D^+$ magnetic moment~%
\cite{Kotulla:2002cg}
has recently appeared, and further data are eagerly awaited.
 
Although more experimental data for the other
decuplet electromagnetic moments
can be expected in the future, 
progress will be slow due to significant experimental difficulties.
Theory, however, may be able to catch up.
While lattice calculations using the quenched approximation have already appeared~\cite{Leinweber:1992hy},
we expect partially quenched calculations for decuplet
observables in the near future.
One problem that currently and foreseeably plagues 
these lattice calculations 
is that they are performed with unphysically large quark masses. 
Therefore, to make physical predictions,
one must extrapolate from the heavier quark masses
used on the lattice 
(currently on the order of the strange quark mass) down to
the physical light quark masses.
For quenched QCD (QQCD) lattice calculations, 
where the fermion determinant that
arises from the path integral is set equal to one,  
one uses quenched chiral perturbation theory (\QCPT)%
~\cite{Morel:1987xk,Sharpe:1992ft,Bernard:1992ep,
Bernard:1992mk,Golterman:1994mk,Sharpe:1996qp,Labrenz:1996jy}
to extrapolate.
The problem with the quenched approximation is
that the Goldstone boson singlet, 
the $\eta'$, which is heavy in QCD,
remains light and must be included in the
\QCPT\ Lagrangian.
This requires new operators and hence new
low-energy constants.
Thus in general, the low-energy constants appearing in \QCPT\
are unrelated to those in \CPT\ and
extrapolated quenched lattice data is unrelated to QCD.
In fact, several examples show that
the behavior of meson loops near the chiral limit is
frequently misrepresented in \QCPT%
~\cite{Booth:1994rr,Kim:1998bz,Savage:2001dy,Arndt:2002ed,Dong:2003im,Arndt:2003ww}.
We find this to be true for the decuplet 
electromagnetic observables. Indeed, to the order we work only
quenched chiral singularities are present in the quenched 
electromagnetic moments; the charge radii have no quark mass
dependence at all.

The unattractive features of QQCD can be remedied by using partially quenched 
lattice QCD (PQQCD).  
Unlike in QQCD, where the ``sea quarks'' are neglected by setting the fermion determinant
to unity, in PQQCD these sea quarks are kept as dynamical degrees of freedom and 
their masses can be varied independently
of the valence quark masses.  
For computational reasons they are usually chosen to be heavier. 
The low-energy effective theory of PQQCD is \PQCPT%
~\cite{Bernard:1994sv,Sharpe:1997by,Golterman:1998st,Sharpe:1999kj,
Sharpe:2000bn,Sharpe:2000bc,Sharpe:2001fh,Shoresh:2001ha}.
Since PQQCD retains a $U(1)_A$ anomaly,
the equivalent to the singlet field in QCD is heavy (on the order
of
the chiral symmetry breaking scale $\L_\chi$) and can be integrated out%
~\cite{Sharpe:2000bn,Sharpe:2001fh}.
Therefore, the low-energy constants appearing in \PQCPT\ are the same
as those appearing in \CPT.
By fitting \PQCPT\ to partially quenched
lattice data one can determine these constants  
and actually make physical predictions for QCD.
\PQCPT\ has been used recently to study 
heavy meson~\cite{Savage:2001jw} and
octet baryon observables%
~\cite{Chen:2001yi,Beane:2002vq,Savage:2002fm,Leinweber:2002qb,Arndt:2003ww}.

While there has been a quenched lattice calculation 
of the decuplet magnetic moments%
~\cite{Leinweber:1992hy},
there are currently no partially quenched simulations.
However, in light of the progress that lattice gauge theory has made recently
in the one-hadron sector and the prospect of simulations  
in the two-hadron sector%
~\cite{LATTICEproposal1,LATTICEproposal2,Beane:2002np,Beane:2002nu,Arndt:2003vx},
we expect to see partially quenched calculations of the
decuplet electromagnetic form factors in the near future.

The paper is organized as follows.
First, in Section~\ref{sec:PQCPT}, 
we review \PQCPT\ including the treatment
of the baryon octet and decuplet in the heavy baryon approximation.
In Section~\ref{sec:ff},
we calculate the electromagnetic moments and charge radii of the 
decuplet baryons in both \QCPT\ and \PQCPT\
up to next-to-leading (NLO) order in the chiral expansion.
We use the heavy baryon formalism of Jenkins and Manohar%
~\cite{Jenkins:1991jv,Jenkins:1991ne},
and work to lowest order in the heavy baryon expansion.
These calculations are done in the 
isospin limit of $SU(3)$ flavor. Expressions for form factors with
the $q^2$ dependence at one loop are given in Appendix \ref{s:q-dep}.  
For completeness we also provide the
\PQCPT\ 
result for the baryon quartet electromagnetic moments and  charge radii 
for the $SU(2)$ chiral Lagrangian with non-degenerate quarks in Appendix \ref{s:su2}.
We conclude in Section~\ref{sec:conclusions}.

\section{\label{sec:PQCPT}\PQCPT}
In PQQCD the quark part of the Lagrangian is written as%
~\cite{Sharpe:2000bn,Sharpe:2001fh,Sharpe:2000bc,Sharpe:1999kj,
Golterman:1998st,Sharpe:1997by,Bernard:1994sv,Shoresh:2001ha}
\begin{eqnarray}\label{eqn:LPQQCD}
  {\cal L}
  &=&
  \sum_{a,b=u,d,s}\bar{q}_a(i\Dslash-m_q)_{ab} q_b
  + \sum_{\tilde{a},\tilde{b}=\tilde{u},\tilde{d},\tilde{s}}
      \bar{\tilde{q}}_{\tilde{a}}
      (i\Dslash-m_{\tilde{q}})_{\tilde{a}\tilde{b}} 
      \tilde{q}_{\tilde{b}}
  +
  \sum_{a,b=j,l,r}
  \bar{q}_{\text{sea},a} (i\Dslash-m_{\text{sea}})_{ab} q_{\text{sea},b}
               \nonumber \\
  &&=
  \sum_{j,k=u,d,s,\tilde{u},\tilde{d},\tilde{s},j,l,r}
  \bar{Q}_j(i\Dslash-m_Q)_{jk} Q_k
.\end{eqnarray}
Here, in addition to the fermionic light valence quarks $u$, $d$, and $s$, 
their bosonic counterparts $\tilde{u}$, $\tilde{d}$, and $\tilde{s}$ (the 
ghost quarks) and 
three light fermionic sea quarks $j$, $l$, and $r$ have been added.
These nine quarks are in the fundamental representation of
the graded group $SU(6|3)$%
~\cite{BahaBalantekin:1981kt,BahaBalantekin:1981qy,BahaBalantekin:1982bk}
and appear in the vector
\begin{equation}
  Q=(u,d,s,j,l,r,\tilde{u},\tilde{d},\tilde{s})
\end{equation}
that obeys the graded equal-time commutation relation
\begin{equation} \label{eqn:commutation}
  Q^\a_i({\bf x}){Q^\b_j}^\dagger({\bf y})
  -(-1)^{\eta_i \eta_j}{Q^\b_j}^\dagger({\bf y})Q^\a_i({\bf x})
  =
  \d^{\a\b}\d_{ij}\d^3({\bf x}-{\bf y})
,\end{equation}
where $\a$ and $\b$ are spin and $i$ and $j$ are flavor indices.
The graded equal-time commutation relations for two $Q$'s and two
$Q^\dagger$'s can be written analogously.
The grading factor 
\begin{equation}
   \eta_k
   = \left\{ 
       \begin{array}{cl}
         1 & \text{for } k=1,2,3,4,5,6 \\
         0 & \text{for } k=7,8,9
       \end{array}
     \right.
\end{equation}
reflects the different statistics for
fermionic and bosonic quarks.
The quark mass matrix is given by 
\begin{equation}
  m_Q=\text{diag}(m_u,m_d,m_s,m_j,m_l,m_r,m_u,m_d,m_s)
\end{equation}
so that diagrams with closed ghost quark loops cancel 
those with valence quarks.
Effects of virtual quark loops are,
however, present due to the finite-mass 
sea quarks. 
In the limit $m_j=m_u$, $m_l=m_d$, and $m_r=m_s$ one recovers
QCD.

The light quark electric charge matrix $\cQ$ is not uniquely
defined in PQQCD~\cite{Golterman:2001yv}.  
The only constraint one imposes is for
the charge matrix $\cQ$ to have vanishing
supertrace, so that no new operators
involving the singlet component are subsequently introduced.
Following the convention in~\cite{Chen:2001yi}, we use
\begin{equation}
  \cQ
  =
  \diag
  \left(
    \frac{2}{3},-\frac{1}{3},-\frac{1}{3},q_j,q_l,q_r,q_j,q_l,q_r
  \right)
.\end{equation}
When $m_j\to m_u$, $m_l\to m_d$, and $m_r\to m_s$,
 QCD is recovered independently of the $q$'s.

\subsection{Mesons}
For massless quarks,
the Lagrangian in Eq.~(\ref{eqn:LPQQCD}) exhibits a graded symmetry
$SU(6|3)_L \otimes SU(6|3)_R \otimes U(1)_V$ that is assumed 
to be spontaneously broken down to $SU(6|3)_V \otimes U(1)_V$. 
The low-energy effective theory of PQQCD that emerges by 
expanding about the physical vacuum state is \PQCPT.
The dynamics of the emerging 80~pseudo-Goldstone mesons 
can be described at lowest 
order in the chiral expansion by the Lagrangian
\begin{equation}\label{eqn:Lchi}
  {\cal L} =
  \frac{f^2}{8}
    \str\left(D^\mu\Sigma^\dagger D_\mu\Sigma\right)
    + \l\,\str\left(m_Q\Sigma+m_Q^\dagger\Sigma^\dagger\right)
    + \a\partial^\mu\Phi_0\partial_\mu\Phi_0
    - \mu_0^2\Phi_0^2
\end{equation}
where
\begin{equation} \label{eqn:Sigma}
  \Sigma=\exp\left(\frac{2i\Phi}{f}\right)
  = \xi^2
,\end{equation}
\begin{equation}
  \Phi=
    \left(
      \begin{array}{cc}
        M & \chi^{\dagger} \\ 
        \chi & \tilde{M}
      \end{array}
    \right)
,\end{equation}
$f=132$~MeV,
and the gauge-covariant derivative is
$D_\mu\S=\partial_\mu\S+ie\cA_\mu[\cQ,\S]$.
The str() denotes a supertrace over flavor indices.
The $M$, $\tilde{M}$, and $\chi$ are matrices
of pseudo-Goldstone bosons with quantum numbers of $q\ol{q}$ pairs,
pseudo-Goldstone bosons with quantum numbers of 
$\tilde{q}\ol{\tilde{q}}$ pairs, 
and pseudo-Goldstone fermions with quantum numbers of $\tilde{q}\ol{q}$ pairs,
respectively.
$\Phi$ is defined in the quark basis and normalized such that
$\Phi_{12}=\pi^+$ (see, for example,~\cite{Chen:2001yi}).
Upon expanding the Lagrangian in \eqref{eqn:Lchi} one finds that
to lowest order
the mesons with quark content $Q\bar{Q'}$
are canonically normalized when
their masses are given by
\begin{equation}\label{eqn:mqq}
  m_{QQ'}^2=\frac{4\lambda}{f^2}(m_Q+m_{Q'})
.\end{equation}

The flavor singlet field given by $\Phi_0=\str(\Phi)/\sqrt{6}$
is, in contrast to the \QCPT\ case, rendered heavy by the $U(1)_A$
anomaly
and can therefore be integrated out in \CPT.
Analogously its mass $\mu_0$ can be taken to be 
on the order of the chiral symmetry breaking scale, 
$\mu_0\to\Lambda_\chi$.  
In this limit the 
flavor singlet propagator becomes independent of the
coupling $\a$ and 
deviates from a simple pole form~\cite{Sharpe:2000bn,Sharpe:2001fh}. 

\subsection{Baryons}
Just as there are mesons in PQQCD with quark content
$\ol{Q}_iQ_j$ that contain 
valence, sea, and ghost quarks, there are baryons 
with quark composition $Q_iQ_jQ_k$ that
contain all three types of quarks.
Restrictions on the baryon fields ${\mathcal B}_{ijk}$
come from the fact that these fields
must reproduce the familiar octet and decuplet
baryons when $i$, $j$, $k=1$-$3$~\cite{Labrenz:1996jy,Chen:2001yi}.
To this end, one decomposes the irreducible representations
of $SU(6|3)_V$ into 
irreducible representations of 
$SU(3)_{\text{val}} \otimes SU(3)_{\text{sea}} \otimes SU(3)_{\text{ghost}}
 \otimes U(1)$.
The method to construct the octet baryons is to use the
interpolating field
\begin{equation}
  \cB_{ijk}^\g
  \sim
  \left(Q_i^{\a,a}Q_j^{\b,b}Q_k^{\g,c}-Q_i^{\a,a}Q_j^{\g,c}Q_k^{\b,b}\right)
  \e_{abc}(C\g_5)_{\a\b}
.\end{equation}
The spin-1/2 baryon octet $B_{ijk}=\cB_{ijk}$,
where the
indices $i$, $j$, and $k$ are restricted to $1$--$3$,
is contained as a $(\bf 8,\bf 1,\bf1)$ of
$SU(3)_{\text{val}} \otimes SU(3)_{\text{sea}} \otimes SU(3)_{\text{ghost}}$
in the $\bf 240$ representation.
The octet baryons, written in the familiar two-index notation
\begin{equation}
  B=
    \left(
      \begin{array}{ccc}
        \frac{1}{\sqrt{6}}\L+\frac{1}{\sqrt{2}}\S^0 & \S^+ & p \\ 
        \S^- & \frac{1}{\sqrt{6}}\L-\frac{1}{\sqrt{2}}\S^0 & n \\
        \Xi^- & \Xi^0 & -\frac{2}{\sqrt{6}}\L
      \end{array}
    \right)
,\end{equation}
are embedded in $B_{ijk}$ as~\cite{Labrenz:1996jy}
\begin{equation}
  B_{ijk}
  =
  \frac{1}{\sqrt{6}}
  \left(
    \e_{ijl}B_{kl}+\e_{ikl}B_{jl}
  \right)
.\end{equation}

Besides the conventional octet baryons that contain valence quarks,
$qqq$,
there are also baryon fields with sea and ghost quarks
contained in the $\bf 240$, 
e.g., $qq_{\text{sea}}\tilde{q}$.  
The baryon states needed for our calculation have at most one ghost or one sea quark
and have been constructed explicitly in~\cite{Chen:2001yi}.

Similarly, 
the familiar spin-3/2 decuplet baryons are embedded
in the $\bf 165$.  
Here,
one uses the interpolating field
\begin{equation} \label{eqn:Tstate}
  \cT_{ijk}^{\a,\mu}
  \sim
  \left(
    Q_i^{\a,a}Q_j^{\b,b}Q_k^{\g,c}
    +Q_i^{\b,b}Q_j^{\g,c}Q_k^{\a,a}
    +Q_i^{\g,c}Q_j^{\a,a}Q_k^{\b,b}
  \right)
  \e_{abc}
  \left(C\g^\mu\right)_{\b\g}
\end{equation}
that describes the $\bf 165$ dimensional representation of $SU(6|3)_V$. 
The decuplet baryons $T_{ijk}$
are then readily embedded in $\cT$ by construction:
$T_{ijk}=\cT_{ijk}$, where
the indices $i$, $j$, and $k$ are restricted to $1$--$3$.
They transform as a $(\bf 10, \bf 1, \bf1)$ under
$SU(3)_{\text{val}} \otimes SU(3)_{\text{sea}} \otimes SU(3)_{\text{ghost}}$.
Because of Eqs.~(\ref{eqn:commutation}) and \eqref{eqn:Tstate}, $T_{ijk}$ is
a totally symmetric tensor.  
Our normalization convention is such that $T_{111}=\D^{++}$.
For the spin-3/2 baryons consisting of two valence and one ghost quark
or two valence and one sea quark, we use the states constructed in%
~\cite{Chen:2001yi}.

At leading order in the heavy baryon expansion, the 
free Lagrangian for the $\cB_{ijk}$ and 
$\cT_{ijk}$ is given by~\cite{Labrenz:1996jy}
\begin{eqnarray} \label{eqn:L}
  {\mathcal L}
  &=&
  i\left(\ol\cB v\cdot{\mathcal D}\cB\right)
  +2\a_M\left(\ol\cB \cB{\mathcal M}_+\right)
  +2\b_M\left(\ol\cB {\mathcal M}_+\cB\right)
  +2\sigma_M\left(\ol\cB\cB\right)\str\left({\mathcal M}_+\right)
                              \nonumber \\
  &&-i\left(\ol\cT^\mu v\cdot{\mathcal D}\cT_\mu\right)
  +\D\left(\ol\cT^\mu\cT_\mu\right)
  +2\g_M\left(\ol\cT^\mu {\mathcal M}_+\cT_\mu\right)
  -2\ol\sigma_M\left(\ol\cT^\mu\cT_\mu\right)\str\left({\mathcal M}_+\right)
,\end{eqnarray}
where 
${\mathcal M}_+
  =\frac{1}{2}\left(\xi^\dagger m_Q \xi^\dagger+\xi m_Q \xi\right)$.
The brackets in (\ref{eqn:L}) are shorthands for field
bilinear invariants originally employed in~\cite{Labrenz:1996jy}.
To lowest order in the chiral expansion, Eq.~\eqref{eqn:L} gives the 
propagators
\begin{equation}
  \frac{i}{v\cdot k},\quad
  \frac{iP^{\mu\nu}}{v\cdot k-\D}
\end{equation}
for the spin-1/2 and spin-3/2 baryons, respectively.
Here, $v$ is the velocity and $k$ the residual momentum of the
heavy baryon which are related to the momentum $p$ by
$p=M_B v+k$.
$M_B$ denotes the (degenerate) mass of the octet baryons
and $\D$ the decuplet--baryon mass splitting.
The polarization tensor
\begin{equation}
  P^{\mu\nu}
  =
  \left(v^\mu v^\nu-g^{\mu\nu}\right)-\frac{4}{3}S_v^\mu S_v^\nu
\end{equation}
reflects the fact that the Rarita-Schwinger field 
$(\cT^\mu)_{ijk}$ contains both spin-1/2 and spin-3/2 pieces;
only the latter remain as propagating
degrees of freedom~\cite{Jenkins:1991ne}.

The Lagrangian describing the relevant interactions of the $\cB_{ijk}$ 
and $\cT_{ijk}$ 
with the pseudo-Goldstone mesons is
\begin{equation} \label{eqn:Linteract}
  {\cal L} =   2{\mathcal H}\left(\ol{\cT}^\nu S^\mu A_\mu \cT_\nu\right) 
 + \sqrt{\frac{3}{2}}\cC
  \left[
    \left(\ol{\cT}^\nu A_\nu \cB\right)+ \text{h.c.}
  \right]  
.\end{equation}
The axial-vector and vector meson fields $A^\mu$ and $V^\mu$
are defined by analogy to those in QCD:
\begin{equation}
  A^\mu=\frac{i}{2}
        \left(
          \xi\partial^\mu\xi^\dagger-\xi^\dagger\partial^\mu\xi
        \right)\quad\text{and}\quad
  V^\mu=\frac{1}{2}
        \left(
          \xi\partial^\mu\xi^\dagger+\xi^\dagger\partial^\mu\xi
        \right)
.\end{equation}
The latter appears in Eq.~\eqref{eqn:Linteract} in the
covariant derivatives of $\cB_{ijk}$ and $\cT_{ijk}$ 
that both have the form
\begin{equation}
  ({\mathcal D}^\mu \cB)_{ijk}
  =
  \partial^\mu \cB_{ijk}
  +(V^\mu)_{il}\cB_{ljk}
  +(-)^{\eta_i(\eta_j+\eta_m)}(V^\mu)_{jm}\cB_{imk}
  +(-)^{(\eta_i+\eta_j)(\eta_k+\eta_n)}(V^\mu)_{kn}\cB_{ijn}
.\end{equation}

\section{Decuplet electromagnetic properties}
\label{sec:ff}
The electromagnetic moments of decuplet baryons in \CPT\ have been investigated previously 
in~\cite{Butler:1994ej,Banerjee:1996wz}. Additionally there has been interest in the decuplet electromagnetic properties
in the large $N_c$ limit of QCD~\cite{Buchmann:2000wf,Buchmann:2002mm,Buchmann:2002et,Cohen:2002sd}.
In this Section we calculate the decuplet electromagnetic moments and charge radii in
\PQCPT\ and \QCPT.
The basic conventions and notations for the
mesons and baryons in \PQCPT\ have been laid forth
in the last Section; \QCPT\ has been extensively reviewed in 
the literature%
~\cite{Morel:1987xk,Sharpe:1992ft,Bernard:1992ep,
Bernard:1992mk,Golterman:1994mk,Sharpe:1996qp,Labrenz:1996jy}.
Additionally the decuplet charge radii in \CPT\ are provided since 
they have not been calculated before. First we review the electromagnetic form
factors of heavy spin-$3/2$ baryons.
 
Using the heavy baryon formalism%
~\cite{Jenkins:1991jv,Jenkins:1991ne}, 
the decuplet matrix elements of the electromagnetic current $J^\rho$ can be parametrized as
\begin{equation}
\langle \ol T(p') | J^\rho | T(p) \rangle = -  \ol u_\mu(p') \mathcal{O}^{\mu \rho \nu} u_\nu(p),
\end{equation}
where $u_\mu(p)$ is a Rarita-Schwinger spinor for an on-shell  heavy baryon satisfying
$v^\mu u_\mu(p) = 0$ and $S^\mu u_\mu(p) = 0$.
The tensor $\mathcal{O}^{\mu \rho \nu}$ can be parametrized in terms of four independent, 
Lorentz invariant form factors
\begin{equation}
\mathcal{O}^{\mu \rho \nu} = g^{\mu \nu} \left\{ v^\rho F_1(q^2) + \frac{[S^\rho,S^\tau] }{M_B}q_\tau F_2(q^2)  \right\}
+ \frac{q^\mu q^\nu}{(2 M_B)^2} \left\{ v^\rho G_1(q^2) + \frac{[S^\rho,S^\tau]}{M_B} q_\tau G_2(q^2)  \right\},
\end{equation}
where the momentum transfer $q = p' - p$. 
The form factor $F_1(q^2)$ is normalized to the decuplet charge in units of $e$: $F_1(0) = Q$.
At NLO in the chiral expansion, the form factor $G_2(q^2) = 0$.

Extraction of the form factors requires a nontrivial identity for on-shell 
Rarita-Schwinger spinors~\cite{Nozawa:1990gt}. For heavy baryon spinors, the 
identity is
\begin{equation}
 \ol u_\a(p')\big( q^\alpha g^{\mu \beta} - q^\beta g^{\mu \alpha} \big) u_\b(p)  =  \ol u_\a(p') 
\left[ - \frac{q^2}{2 M_B} g^{\alpha \beta} v^\mu  + 2  g^{\alpha \beta} [S^\mu,S^\nu]q_\nu 
+ \frac{1}{M_B} q^\alpha q^\beta v^\mu \right] u_\b(p) .
\end{equation}
Linear combinations of the above (Dirac- and Pauli-like) form 
factors make the electric charge $G_{E0}(q^2)$, magnetic dipole $G_{M1}(q^2)$, 
electric quadrupole
$G_{E2}(q^2)$, and magnetic octupole $G_{M3}(q^2)$ form factors. 
This conversion from covariant vertex functions to multipole form 
factors for spin-$3/2$ particles is explicated in~\cite{Nozawa:1990gt}.
For our calculations, the charge radius
\begin{equation} \label{eqn:fred}
< r_E^2 > \equiv 6 \frac{ d}{dq^2}G_{E0}(q^2) \Bigg|_{q^2 = 0} 
= 
6 \left\{ \frac{d F_1 (0) }{dq^2} - 
\frac{1}{12 M_B^2} \left[2 Q - 3 F_2(0) - G_1(0) \right]\right\}, 
\end{equation}
the magnetic moment 
\begin{equation}
\mu \equiv G_{M1}(0) - Q = F_{2}(0),
\end{equation}
and the electric quadrupole moment 
\begin{equation}
\mathbb{Q} \equiv G_{E2}(0) - Q = -\frac{1}{2} G_1(0).
\end{equation}
To the order we work in the chiral expansion, the magnetic octupole moment is zero.

\subsection{\PQCPT}
Let us first consider the calculation of the decuplet baryon electromagnetic properties
in \PQCPT.
Here, the leading tree-level contributions to the magnetic moments 
come  from the dimension-5 operator%
\footnote{We use
$F_{\mu\nu}=\partial_\mu A_\nu-\partial_\nu A_\mu$.}
\begin{equation} \label{eqn:LDF}
\cL = \mu_c \frac{ 3 i e }{M_B}  \big(\ol\cT_\mu \cQ \cT_\nu \big) F^{\mu \nu}, 
\end{equation}
which matches onto the \CPT\ operator~\cite{Jenkins:1993pi}
\begin{equation}
\cL = \mu_c \frac{i e  Q_{i}}{M_B} \ol T{}^{i}_\mu T^{i}_\nu F^{\mu \nu},
\end{equation}
when the indices in Eq.~\eqref{eqn:LDF} are restricted to $1$--$3$. Here $Q_{i}$ is the charge of the $i$th decuplet state.
Additional tree-level contributions come from the leading dimension-6 electric quadrupole operator
\begin{equation} \label{eqn:Lc}
\cL = - \mathbb{Q}_{\text{c}} \frac{3 e}{\L_\chi^2} \big(\ol \cT{}^{\{\mu} \cQ \cT^{\nu\}} \big)  v^\alpha \partial_\mu F_{\nu \alpha}.
\end{equation}
Here the action of ${}^{\{}\ldots{}^{\}}$ on Lorentz indices produces the symmetric traceless part of the tensor, 
{\it viz.}, $\mathcal{O}^{\{\mu \nu\}} = \mathcal{O}^{\mu \nu} + \mathcal{O}^{\nu \mu} - 
\frac{1}{2} g^{\mu\nu} \mathcal{O}^{\alpha}{}_{\alpha}$. The operator in Eq.~\eqref{eqn:Lc}
matches onto the \CPT\ operator~\cite{Butler:1994ej}
\begin{equation}
\cL = - \mathbb{Q}_{\text{c}} \frac{e Q_i}{\L_\chi^2} \ol T{}^{\{\mu}_{i} T^{\nu\}}_{i} v^\alpha \partial_\mu F_{\nu \alpha} .
\end{equation}
The final tree-level contributions arise from the leading dimension-6 charge radius operator
\begin{equation} \label{eqn:Lnew}
\cL =  c_c \frac{3 e}{\L_\chi^2} \big( \ol \cT{}^\sigma \cQ \cT_{\sigma} \big) v_\mu \partial_\nu F^{\mu \nu}
\end{equation}
which matches onto the \CPT\ operator
\begin{equation}
\cL = c_c \frac{e  Q_{i}}{\L_\chi^2}  \ol T{}^\sigma_{i} T_{\sigma,i} \,  v_\mu \partial_\nu F^{\mu \nu}.
\end{equation}
Notice that the PQQCD low-energy constants $\mu_c$, $\mathbb{Q}_{\text{c}}$,
and $c_c$ have the same numerical values as in QCD.

\begin{figure}
\caption{Loop diagrams contributing to the decuplet electromagnetic moments and charge radii. A thick (thin)
solid line denotes a decuplet (octet) baryon. The diagrams in the first row contribute to the electromagnetic
moments and charge radii. The remaining diagrams with a photon have no $q^2$-dependence. These, along with the
wavefunction renormalization diagrams, ensure non-renormalization of the electric charge.}
\epsfig{file=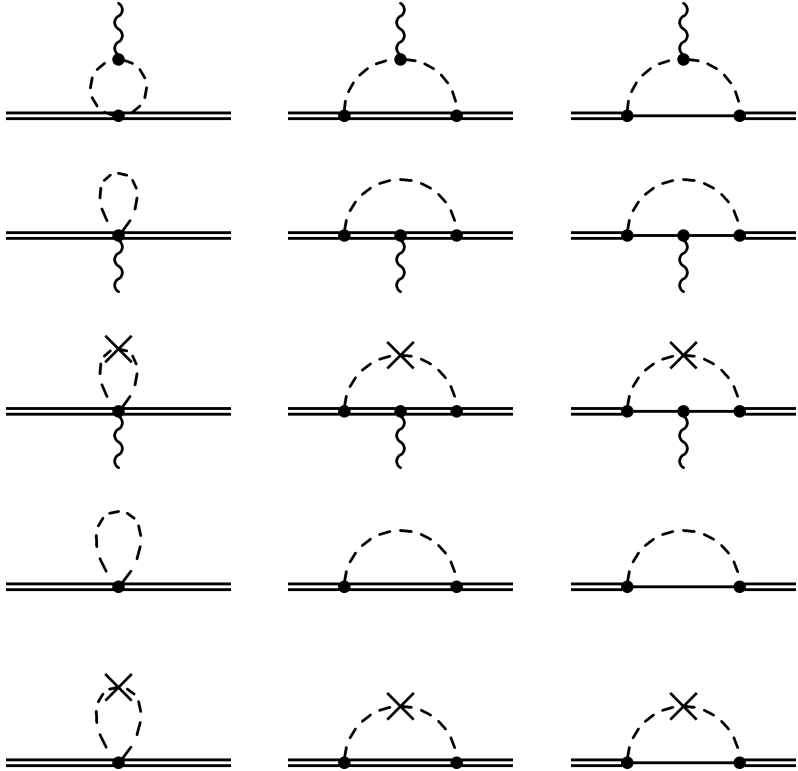}
\label{F:Fdecuplet}
\end{figure}

The NLO contributions to electromagnetic observables in the chiral expansion arise from the one-loop diagrams 
shown in Fig. \ref{F:Fdecuplet}. Calculation of these diagrams yields
\begin{multline}
F_1(q^2)  = Q \left( 1 - \frac{\mu_c q^2}{2 M_B^2} - \frac{ \mathbb{Q}_{\text{c}} q^2}{2 \L_\chi^2} + \frac{c_c q^2}{\L_\chi^2} \right)
-\frac{1}{6} q^2 \frac{3 + \mathcal{C}^2}{16 \pi^2 f^2}  \sum_X  A_X  \log \frac{m_X^2}{\mu^2} \\
 - \frac{11}{54} q^2  \frac{\mathcal{H}^2}{16 \pi^2 f^2}  \sum_X A_X \left[ \log \frac{m_X^2}{\mu^2} - 
\frac{\Delta m_X}{\Delta^2 - m^2_X}  \cR \left(\frac{\Delta}{m_X}\right) \right] + \mathcal{O}(q^4), 
\end{multline}
\begin{equation} \label{eqn:F2}
F_2(0)  = \mu = 2 \mu_c Q + \frac{ M_B \mathcal{H}^2}{36 \pi^2 f^2} \sum_X A_X \left[ \Delta \log \frac{m_X^2}{\mu^2} - 
m_X \cR \left(\frac{\Delta}{m_X} \right)  \right] - 
\frac{\mathcal{C}^2 M_B}{8 \pi f^2} \sum_X A_X m_X, 
\end{equation}
and
\begin{multline} \label{eqn:G1}
G_1(0) = - 2 \mathbb{Q} =  4 Q \left( \mu_c + \mathbb{Q}_c \frac{2 M_B^2}{\L_\chi^2}  \right) - \frac{M_B^2 \mathcal{C}^2}{12 \pi^2 f^2} 
\sum_X A_X \log \frac{m_X^2}{\mu^2}  \\ 
+ \frac{M_B^2 \mathcal{H}^2}{27 \pi^2 f^2} \sum_X A_X \left[ \log \frac{m_X^2}{\mu^2} - \frac{\Delta m_X}{\Delta^2 - m^2_X}  \cR \left( \frac{\Delta}{m_X} \right) \right].
\end{multline}
The only loop contributions kept in the above expressions are those non-analytic in the quark masses. The full $q^2$ dependence
at one-loop has been omitted from the above expressions but is given in Appendix \ref{s:q-dep}.
The function $\mathcal{R}(x)$ is given by
\begin{equation}
\cR (x) = \sqrt{x^2 - 1} \, \log \frac{x - \sqrt{x^2 - 1 + i \epsilon}}{x + \sqrt{x^2 - 1 + i \epsilon}}
.\end{equation}
The sum in the above expressions is over all possible loop mesons $X$.
The computed values of the coefficients $A_X^T$ that appear above are listed in Table \ref{t:clebsch} for 
each of the decuplet states $T$.  
In the table we have listed values corresponding to the loop meson that has mass $m_X$ for both \CPT\
and \PQCPT.  In particular, the \CPT\ coefficients can be used to find the QCD decuplet charge radii, 
which have not been calculated before. Using Eq.~\eqref{eqn:fred}, the expression for the charge radii is 

\begin{table}
\caption{The coefficients $A_X^T$ for SU($3$) for each of the decuplet states in \CPT\ and \PQCPT.
The index $X$ corresponds to the loop meson that has mass $m_X$.}  
\begin{ruledtabular}
\begin{tabular}{l | c c | c  c  c  c  c  c  c }

	    & \multicolumn{2}{c |}{\CPT}  &  \multicolumn{7}{c}{\PQCPT} \\
	              & $\pi$ & $K$  &   $\pi$   &   $K$   &  $\eta_s$  &   $ju$   &   $ru$   & $js$   & $rs$ \\
	 	
	\hline
	$\Delta^{++}$       & $1$ & $1$  & $-\frac{1}{3} + q_{jl}$ & $\frac{1}{3} + q_r$ & $0$ & $\frac{4}{3} - q_{jl}$ & 
												$\frac{2}{3} - q_{r}$ & $0$ & $0$ \\
 
	$\Delta^{+}$        & $\frac{1}{3}$ & $\frac{2}{3}$ & `` & `` & `` & 
										$\frac{2}{3} - q_{jl}$ & $\frac{1}{3} - q_r$ & `` & `` \\

	$\Delta^{0}$        & $-\frac{1}{3}$ & $\frac{1}{3}$  & `` & `` & `` & $-q_{jl}$ & 
														$-q_r$ & `` & `` \\

	$\Delta^{-}$        & $-1$ & $0$ & `` & `` & `` & $-\frac{2}{3} - q_{jl}$ & 
													$-\frac{1}{3} - q_r$ & `` & `` \\

	$\Sigma^{*,+}$       & $\frac{2}{3}$ & $\frac{1}{3}$  & $\;- \frac{2}{9} + \frac{2}{3} q_{jl}\;$ & 
$\;\frac{1}{9} + \frac{2}{3} q_r + \frac{1}{3} q_{jl}\;$ & $\;\frac{1}{9}  + \frac{1}{3} q_r\;$ & $\;\frac{8}{9}  - \frac{2}{3} q_{jl}\;$ 
& $\;\frac{4}{9} - \frac{2}{3} q_r \;$ & $\;-\frac{2}{9} - \frac{1}{3} q_{jl}\;$ &  $\;-\frac{1}{9} -\frac{1}{3} q_r$ \\

	$\Sigma^{*,0}$       & $0$ & $0$  & `` & `` & ``  & $\frac{2}{9} - \frac{2}{3} q_{jl}$ & $\frac{1}{9} - \frac{2}{3} q_r$ & ``
 & `` \\	

	$\Sigma^{*,-}$      & $-\frac{2}{3}$ & $-\frac{1}{3}$  & ``  & `` & `` & $- \frac{4}{9} - \frac{2}{ 3} q_{jl}$ & $ - \frac{2}{9} -\frac{2}{ 3} q_r$ 
& `` & `` \\

	$\Xi^{*,0}$         & $\frac{1}{3}$ & $-\frac{1}{3}$  & $-\frac{1}{9} + \frac{1}{3} q_{jl}$ & 
$- \frac{1}{9}+ \frac{1}{ 3} q_{r}+ \frac{2}{3} q_{jl}$ 
& $\frac{2}{9} + \frac{2}{3} q_{r}$ & $\frac{4}{9}- \frac{1}{3} q_{jl}$ & $\frac{2}{9} - \frac{1}{3} q_r$ & 
$-\frac{4}{9}-\frac{2}{3}q_{jl}$ & $-\frac{2}{9} - \frac{2}{3} q_r$ \\

	$\Xi^{*,-}$          & $-\frac{1}{3}$ & $-\frac{2}{3}$  & `` & `` & `` & $-\frac{2}{9}- \frac{1}{3} q_{jl}$ & $-\frac{1}{9} - \frac{1}{3} q_r$ & ``  & `` \\

	$\Omega^-$          & $0$ & $-1$   & $0$ & $-\frac{1}{3} + q_{jl}$ & $\frac{1}{3} +q_{r}$ & $0$ & $0$ & $-\frac{2}{3} - q_{jl}$ 
& $-\frac{1}{3} -  q_{r}$ \\

\end{tabular}
\end{ruledtabular}
\label{t:clebsch}
\end{table} 

\begin{multline} \label{eqn:r_E}
<r_E^2> = Q \left( \frac{2 \mu_c - 1}{M_B^2} + \frac{\mathbb{Q}_c + 6 c_c}{\L_\chi^2}  \right)
- \frac{1}{3} \, \frac{9 + 5 \mathcal{C}^2}{16 \pi^2 f^2}  \sum_X  A_X  \log \frac{m_X^2}{\mu^2} \\
 - \frac{25}{27}\, \frac{\mathcal{H}^2}{16 \pi^2 f^2} \sum_X A_X \left[ \log \frac{m_X^2}{\mu^2} - 
\frac{\Delta m_X}{\Delta^2 - m^2_X}  \cR \left( \frac{\Delta}{m_X} \right) \right]. 
\end{multline}

\begin{figure}[tb]
\includegraphics[width=0.8\textwidth]{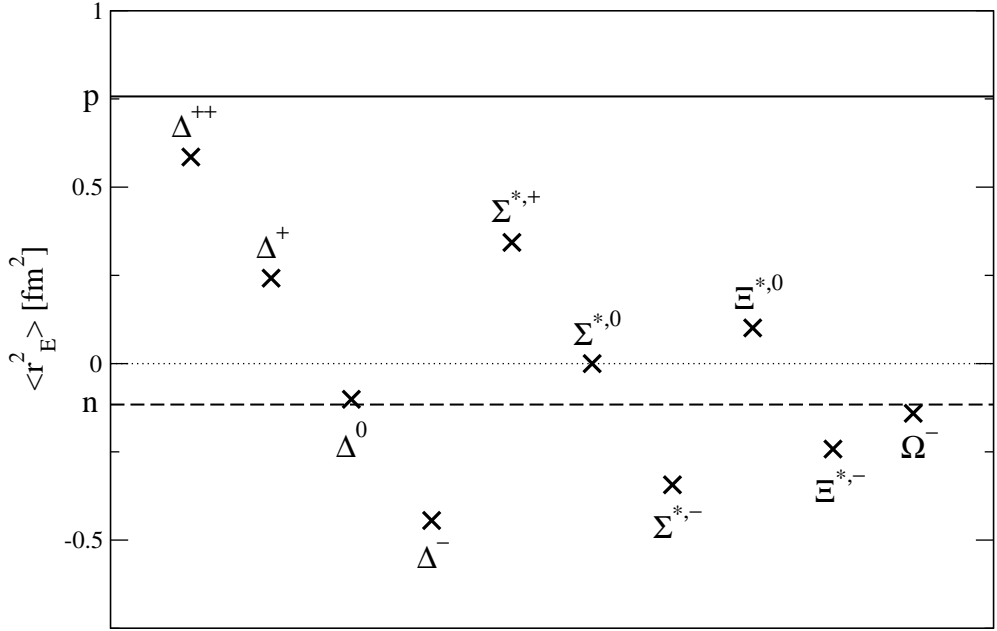}%
\caption{The charge radii of the decuplet baryons in \CPT. The contribution from counterterms has been set to zero. The radii 
(squared) here come from the one-loop diagrams only  and are plotted in units of fm${}^2$. For reference we have shown both the proton and
neutron charge radii (solid and dashed lines, respectively).}
\label{F:radii}
\end{figure}

In the absence of experimental and lattice data for the low energy constants $\mathbb{Q}_c$ and $c_c$, we cannot 
ascertain the contributions to the charge radii from local counterterms in \CPT. 
We can consider, however, just the formally dominant loop contributions.
To this end, we choose the values $\mathcal{C} = - 2 D$ and $\mathcal{H} = - 3 D$ with $D = 0.76$~\cite{Jenkins:1991ne}, 
and the masses
$\D = 270$~MeV, $m_\pi = 140$~MeV, and $m_K = 500$~MeV. The loop contributions to the charge radii in \CPT\ are 
then evaluated for the decuplet at the scale $\mu = 1$~GeV and plotted in Fig.~\ref{F:radii}.  

\subsection{\QCPT}
The calculation of the charge radii and electromagnetic moments can be correspondingly 
executed for \QCPT.
The operators in Eqs.~\eqref{eqn:LDF}, \eqref{eqn:Lc} and \eqref{eqn:Lnew}
contribute, however, their low-energy coefficients
cannot be matched onto QCD.  Therefore we annotate them with a ``Q''.    
The loop contributions encountered in \CPT\ and \PQCPT\ above no longer contribute because  in \QCPT\ $A^T_X = 0$
for all decuplet states $T$. This can be readily seen in two ways. The quenched limit\footnote{In this case, the quenched limit
simply means to remove sea quarks and to fix the charges of the ghost quarks to equal those of their light quark counterparts.}
 of the coefficients in Table \ref{t:clebsch}
makes immediate the vanishing of $A_X^T$. Alternately one can consider the relevant quark flow diagrams with only valence 
quarks in loops. Due to the symmetric nature of $T^{ijk}$, these loops are completely canceled by their ghostly counterparts.
For the charge radii, there are no additional diagrams to consider at this order from singlet hairpin interactions. Thus in
\QCPT, the charge radii have the form
\begin{equation}
<r_E^2> = Q \left( \frac{2 \mu_c^Q - 1}{M_B^2} + \frac{\mathbb{Q}_c^Q + 6 c_c^Q}{\L_\chi^2}  \right),
\end{equation} 
where the dependence on the quark mass enters at the next order.

\begin{figure}
\caption{Hairpin diagrams that give contributions of the form $\sim \mu_o^2 \log m_q$ to decuplet electromagnetic moments
in Q$\chi$PT. The square at the photon vertex represents the relevant operators from Eqs.~\eqref{eqn:LDF}, \eqref{eqn:Lc}, and
\eqref{eqn:Lbob}.}
\epsfig{file=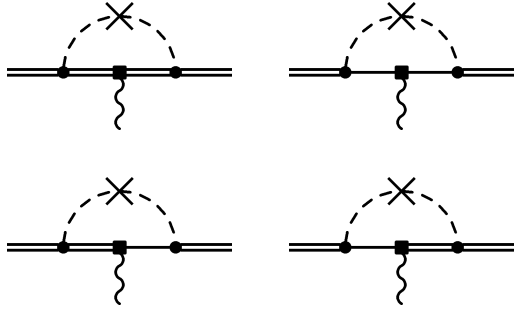}
\label{F:Fquenched}
\end{figure}

Additional terms of the form $\mu_0^2\log m_q$ involving hairpins~\cite{Labrenz:1996jy,Savage:2001dy}
do contribute to the electromagnetic moments as 
they are of the same order in the chiral expansion. 
As explained in~\cite{Chow:1998xc}, the axial hairpins do not contribute.
In the diagrams shown in Fig. \ref{F:Fquenched}, one sees that there are contributions from the 
electromagnetic moments of the decuplet and octet baryons as well as their transition moments.
These interactions are described by the operators in Eqs.~\eqref{eqn:LDF} and \eqref{eqn:Lc} (now
with quenched coefficients) along with
\begin{eqnarray} \label{eqn:Lbob}
\cL 
&=& 
\frac{i e }{2 M_B} 
\left\{\mu_\a^Q \left( \ol \cB [S^\mu,S^\nu] \cB \cQ \right) 
+ \mu_\b^Q \left( \ol \cB [S^\mu,S^\nu] \cQ  \cB \right)    \right\} F_{\mu \nu} \nonumber \\
&&+ \sqrt{\frac{3}{2}}  \mu_T^Q \frac{i e}{2 M_B} \left[ \left( \ol \cB  S^\mu \cQ \cT^\nu \right) + \text{h.c.}  \right] F_{\mu \nu} 
+  \sqrt{\frac{3}{2}}\mathbb{Q}_T^Q \frac{e}{\L_\chi^2} 
\left[ \left( \ol \cB  S^{\{\mu} \cQ \cT^{\nu\}} \right) + \text{h.c.}  \right]  v^\a \partial_\mu F_{\nu \a}. \nonumber \\
\end{eqnarray}
It is easier to work with the combinations $\mu_D^Q$ and $\mu_F^Q$ defined by
\begin{equation}
\mu_\a^Q = \frac{2}{3} \mu_D^Q + 2 \mu_F^Q \quad \text{and} \;\; \mu_\b^Q = -\frac{5}{3} \mu_D^Q + \mu_F^Q.
\end{equation}

To calculate the quenched electromagnetic moments, we also need the hairpin wavefunction renormalization diagrams 
shown in Fig.~\ref{F:Fdecuplet}. These along with the diagrams in Fig.~\ref{F:Fquenched} are economically expressed 
in terms of the function
\begin{align} \label{eqn:I}
I(m_1,m_2,\D_1,\D_2,\mu) & = \frac{Y(m_1,\D_1,\mu) +Y(m_2,\D_2,\mu) - Y(m_1,\D_2,\mu) - Y(m_2,\D_1,\mu)}{(m_1^2 - m_2^2)(\D_1 - \D_2)} 
\notag  \\
& = - i \frac{16 \pi^2}{3} \int \frac{d^D k}{(2\pi)^D} \frac{k^2 - (k\cdot v)^2}{(k^2 - m_1^2)(k^2 - m_2^2) 
(k\cdot v - \D_1) (k\cdot v - \D_2)},
\end{align}
where
\begin{equation}
Y(m,\D,\mu) = \D \left(m^2 - \frac{2}{3} \D^2\right) \log \frac{m^2}{\mu^2} + \frac{2}{3} m (\D^2  - m^2) 
\mathcal{R}\left(\frac{\D}{m}\right)
\end{equation}
and we have kept only non-analytic contributions.  The following shorthands are convenient
\begin{align}
I_{\eta_q \eta_{q^\prime}} & = I(m_{\eta_q},m_{\eta_{q^\prime}},0,0,\mu), \notag \\
I^{\D }_{\eta_q \eta_{q^\prime}} & = I(m_{\eta_q},m_{\eta_{q^\prime}},\D,0,\mu), \notag \\ 
I^{\D \D}_{\eta_q \eta_{q^\prime}} & = I(m_{\eta_q},m_{\eta_{q^\prime}},\D,\D,\mu)
.\end{align}
Specific limits of the function $I_{\eta_q \eta_{q^\prime}}$
appear in~\cite{Savage:2001dy}.
The wavefunction renormalization factors arising from hairpin diagrams can then be expressed as
\begin{equation} \label{eqn:Z}
Z = 1 - \frac{\mu_o^2}{16 \pi^2 f^2} \sum_{XX^\prime} \left[  \frac{1}{2} (\mathcal{C}^Q)^2 C_{XX^\prime} I_{XX^\prime} + \frac{5}{9} (\mathcal{H}^Q)^2 B_{XX^\prime} I^{\D\D}_{XX^\prime} 
 \right].
\end{equation}
\begin{table}
\caption{The SU$(3)$ coefficients $B_{XX^\prime}$ and $C_{XX^\prime}$ in \QCPT.}  
\begin{tabular}{c | c  c  c |  c  c  c }

	    & \multicolumn{3}{c}{$B_{XX^\prime}$}  &\multicolumn{3}{c}{$C_{XX^\prime}$}   \\
	                 & $\quad \eta_u \eta_u$   &   $\quad \eta_u \eta_s$   &   $\quad \eta_s \eta_s$   &  $\quad \eta_u \eta_u$  &   $\quad \eta_u \eta_s$   &   $ \quad \eta_s \eta_s$   \\
	 	
	\hline
	$\Delta^{++},\Delta^{+},\Delta^{0},\Delta^{-} $       &  $1$ & $0$ & $0$   &  $0$ & $0$ & $0$  \\
	$\Sigma^{*,+},\Sigma^{*,0},\Sigma^{*,-}$       	      &  $\frac{4}{9}$ & $\frac{4}{9}$ & $\frac{1}{9}$   &  $\frac{2}{9}$ & $-\frac{4}{9}$ & $\frac{2}{9}$  \\
	$\Xi^{*,0},\Xi^{*,-}$       			      &  $\frac{1}{9}$ & $\frac{4}{9}$ & $\frac{4}{9}$   &  $\frac{2}{9}$ & $-\frac{4}{9}$ & $\frac{2}{9}$  \\
	$\Omega^-$                                            &  $0$ & $0$ & $1$   &  $0$ & $0$ & $0$  \\
\end{tabular}
\label{t:quenched}
\end{table} 
The coefficients $B_{XX^\prime}$ and $C_{XX^\prime}$ are listed in Table \ref{t:quenched}. The sum in Eq.~\eqref{eqn:Z} is
over $XX^\prime = \eta_u \eta_u$, $\eta_u \eta_s$, $\eta_s \eta_s$. Combining these factors with the tree-level contributions
and including the corrections from the diagrams in Fig.~\ref{F:Fquenched}, we arrive at the quenched decuplet magnetic moment
\begin{eqnarray} \label{eqn:qF2}
\mu &=& 2 \mu_c^Q Q Z + \frac{\mu_o^2}{16 \pi^2 f^2} \sum_{XX^\prime} \left[ \frac{1}{2} 
(Q \mu_F^Q + \a \mu_D^Q) (\mathcal{C}^Q)^2 C_{XX^\prime} I_{XX^\prime} \right] \nonumber \\
&&+  \frac{\mu_o^2}{16 \pi^2 f^2} \sum_{XX^\prime} \left[  \frac{22}{27} (\mathcal{H}^Q)^2  \mu_c^Q B_{XX^\prime} Q I^{\D \D}_{XX^\prime} - \frac{2}{9} \mathcal{C}^Q \mathcal{H}^Q \mu_T^Q D_{XX^\prime} I^\D_{XX^\prime} 
\right]
\end{eqnarray}
and the quenched electric quadrupole moment
\begin{eqnarray} \label{eqn:qG1}
\mathbb{Q} &=& - 2  Q \left( \mu_c^Q + \mathbb{Q}_c^Q \frac{2 M_B^2}{\L_\chi^2} \right) Z  - \frac{\mu_o^2}{16 \pi^2 f^2}  
\frac{M_B^2}{\L_\chi^2} \sum_{XX^\prime} 
\left( \frac{8}{3} \mathcal{C}^Q \mathcal{H}^Q \mathbb{Q}_T^Q D_{XX^\prime} I^\D_{XX^\prime}\right) \nonumber \\
&&- \frac{\mu_o^2}{16 \pi^2 f^2}  \sum_{XX^\prime} \left[  (\mathcal{H}^Q)^2 \left( \frac{2}{9} \mu_c^Q + \frac{4}{9} \mathbb{Q}_c^Q \frac{M_B^2}{\L_\chi^2}\right) Q B_{XX^\prime} 
I^{\D \D}_{XX^\prime} - \frac{2}{3} \mathcal{C}^Q \mathcal{H}^Q \mu_T^Q D_{XX^\prime} I^\D_{XX^\prime} 
\right]. \nonumber \\
\end{eqnarray}
In Eq.~\eqref{eqn:qF2} the required values for the constant $\a$ are:  
$\a_{\Sigma^{*,+}},\a_{\Sigma^{*,0}},\a_{\Sigma^{*,-} },\a_{\Xi^{*,-}} = 
\frac{1}{3}$ and $\a_{\Xi^{*,0}} = - \frac{2}{3}$. The coefficients $D_{XX^\prime}$ are listed in Table
\ref{T:D}. If a particular decuplet state is not listed, the value of $D_{XX^\prime}$ is zero for all singlet pairs $XX^\prime$.

\begin{table}
\caption{The SU$(3)$ coefficients $D_{XX^\prime}$ in \QCPT. Decuplet states not listed have $D_{XX'} = 0$.}  
\begin{tabular}{l | c  c  c  }

	  & \multicolumn{3}{c}{$D_{XX^\prime}$}     \\
	                 & $\quad \eta_u \eta_u$   &   $\quad \eta_u \eta_s$   &   $\quad \eta_s \eta_s$    \\
	 	
	\hline
	$\Sigma^{*,+}$  &  $\frac{2}{9}$ & $-\frac{1}{9}$ & $-\frac{1}{9}$ \\ 
	$\Sigma^{*,0}$  &  $\frac{1}{9}$ & $- \frac{1}{18}$ & $-\frac{1}{18}$ \\      
    	$\Xi^{*,0}$     &  $\frac{1}{9}$ & $\frac{1}{9}$ & $-\frac{2}{9}$ \\   			       
\end{tabular}
\label{T:D}
\end{table} 

The above expressions can be used to properly extrapolate quenched lattice data to the physical pion mass.
For example, the expression for the quenched magnetic moments for the $\D$ baryons [Eq.~\eqref{eqn:qF2}]
reduces to
\begin{equation}
\mu = 2 Q \mu_c^Q \left( 1 - \frac{4}{27} \, \frac{\mu_o^2}{16 \pi^2 f^2} (\mathcal{H}^Q)^2 I^{\D \D}_{\eta_u \eta_u}  \right)
.\end{equation}
In the above expression we need 
\begin{equation}
  I^{\D \D}_{XX}
  =
  \log\frac{m_X^2}{\mu^2} - \frac{\D m_X}{\D^2-m_X^2}\cR\left(\frac{\D}{m_X}\right)+\dots   
.\end{equation}
where the ellipses denote terms analytic in $m_X$.
Utilizing a least squares analysis
and using the values $\mu=1$~GeV and $\D=270$~MeV 
, we extrapolate the quenched lattice data~\cite{Leinweber:1992hy} 
to the physical pion mass and find for the $\D$ resonances
\begin{equation}
\mu = 2.89 \, Q \; [\mu_N]
.\end{equation}
This is in contrast to the value $\mu \approx 2.49 \, Q \; [\mu_N]$ found from carrying out a \CPT-type extrapolation~\cite{Cloet:2003jm}.
Notice that for many of the decuplet states, in particular the $\D$ baryons, the quenched magnetic moments and electric
quadrupole moments are proportional to the charge $Q$ (unlike the \CPT\ and \PQCPT\ results). 
This elucidates the trends seen in the quenched lattice data~\cite{Leinweber:1992hy}.

\section{\label{sec:conclusions}Conclusions}
We have calculated the electromagnetic moments and charge radii for the SU($3$) decuplet 
baryons in the isospin limit of \PQCPT\
and also derived the result for
the baryon quartet away from the isospin limit for the $SU(2)$
chiral Lagrangian. The $q^2$ dependence of decuplet form factors at one-loop appears in Appendix \ref{s:q-dep}. 
We have also calculated the \QCPT\ results.

Knowledge of the low-energy behavior of QQCD and PQQCD is crucial
to properly extrapolate lattice calculations from 
the quark masses used on the lattice to those in nature.
For the quenched approximation,
where the quark determinant is replaced by unity,
one uses \QCPT\ to do this extrapolation.  
QQCD, however, has no known
connection to QCD. 
Observables calculated in QQCD are often
found to be more divergent in the chiral limit than those in QCD.
This behavior is due to new operators included
in the QQCD Lagrangian, which are non-existent in QCD.
For the decuplet baryons' electromagnetic moments and charge radii
such operators enter at NNLO in the chiral expansion.
Hence, formally 
our NLO result is not more divergent than its QCD counterpart.
This, however, does not mean that our result is free of quenching
artifacts.
While the expansions about the chiral limit for QCD and QQCD
charge radii are formally similar,
$<r^2>\sim\a+\b\log m_Q+\dots$,
the QQCD result consists \emph{entirely} of quenched oddities:
for all decuplet baryons,
diagrams that have bosonic or fermionic mesons running in loops 
completely cancel so that $\b=0$.  In other words, the quenched decuplet 
charge radii have the behavior $<r^2>\sim\a+\dots$ 
and the result is actually independent of $m_Q$ at this order.
For the quenched decuplet magnetic moments and electric quadrupole moments, 
expansions about the chiral limit are again formally similar:
$\mu \sim \a + \b \log m_Q + \gamma \sqrt{m_Q} + \ldots$ and
$\mathbb{Q} \sim \a + \b \log m_Q + \ldots$. 
However, quenching forces $\gamma = 0$ and both $\b$'s arise from singlet contributions
involving the parameter $\mu_o$, which is of course absent in QCD. Thus
the leading non-analytic quark mass dependence that remains for these observables
is entirely a quenched peculiarity.

PQQCD, on the other hand, is free of such eccentric behavior.
The formal behavior of the electromagnetic observables in the chiral limit
has the same form as in QCD.
Moreover, there is 
a well-defined connection to QCD and one can
reliably extrapolate lattice results down to the 
quark masses of reality. 
The low-energy constants appearing in
PQQCD are the same as those in QCD and
by fitting them in \PQCPT\ one can make predictions for QCD.
Our \PQCPT\ result will 
enable the proper extrapolation of PQQCD lattice simulations
of the decuplet electromagnetic moments and charge radii
and we hope it encourages such simulations in the future.

\begin{acknowledgments}
We thank Martin Savage
for very helpful discussions and for useful comments on the manuscript.
This work is supported in part by the U.S.\ Department of Energy
under Grant No.\ DE-FG03-97ER4014. 
\end{acknowledgments}

\appendix 

\section{\label{s:q-dep} The $q^2$ dependence of decuplet form factors }
For reference, we provide the $q^2$ dependence of the decuplet electromagnetic form factors
defined in Section~\ref{sec:ff} at one-loop order in the chiral expansion. To do so we 
define the function 
\begin{equation}
P_X(x,q^2) = \sqrt{1 - \frac{x (1-x) q^2}{m_X^2}}
.\end{equation}
Then we have
\begin{eqnarray} \label{eqn:F1q}
F_1(q^2)  
&=&
 Q \left( 1 - \frac{\mu_c q^2}{2 M_B^2} - \frac{ \mathbb{Q}_{\text{c}} q^2}{2 \L_\chi^2} + \frac{c_c q^2}{\L_\chi^2} \right) \nonumber \\
&&- \frac{3 + \mathcal{C}^2}{16 \pi^2 f^2}  \sum_X  A_X  \left[ \frac{q^2}{6} \log \frac{m_X^2}{\mu^2} 
- 2 m_X^2 \int_0^1 dx \; P^2_X(x,q^2) \log P_X(x,q^2) \right] \nonumber \\
&&- \frac{\mathcal{H}^2}{24 \pi^2 f^2}  \sum_X A_X \Bigg\{ \frac{11}{36} q^2 \log \frac{m_X^2}{\mu^2} 
+ \frac{5}{3} \D m_X  \cR \left( \frac{\D}{m_X} \right) \nonumber\\
&&\phantom{GWB}-\int_0^1 dx \Bigg[\frac{10}{3} 
               \left( \frac{m_X^2}{2} - \D^2 - \frac{11}{10} x(1-x) q^2 \right) \log P_X(x,q^2) \nonumber \\
&&\phantom{an idiot}+ \D m_X  P_X(x,q^2) \left( \frac{5}{3} + \frac{x(1-x) q^2}{\D^2 - m_X^2 P^2_X(x,q^2)} \right) 
\cR \left(\frac{\Delta}{m_X P_X(x,q^2) }\right) \Bigg]  \Bigg\},\nonumber \\  
\end{eqnarray}
\begin{eqnarray} \label{eqn:F2q}
F_2(q^2)  
&=& 
2 \mu_c Q -\frac{\mathcal{C}^2 M_B}{8 \pi f^2} \sum_X A_X  m_X  \int_0^1 dx \; P_X(x,q^2)
+ \frac{ M_B \mathcal{H}^2}{36 \pi^2 f^2} \sum_X A_X 
\Bigg\{ \Delta \log \frac{m_X^2}{\mu^2} \nonumber \\ 
&&\phantom{indeed}+  \int_0^1 dx \Bigg[ 2 \D \log P_X(x,q^2)
-  m_X P_X(x,q^2)  \cR \left(\frac{\Delta}{m_X P_X(x,q^2) } \right) \Bigg] \Bigg\} 
,\end{eqnarray}
and
\begin{eqnarray} \label{eqn:G1q}
G_1(q^2) 
&=&  
4 Q \left( \mu_c + \mathbb{Q}_c \frac{2 M_B^2}{\L_\chi^2}  \right) 
- \frac{M_B^2 \mathcal{C}^2}{2 \pi^2 f^2} \sum_X A_X \left[ \frac{1}{6} \log \frac{m_X^2}{\mu^2} 
+  \int_0^1 dx \, 2 x(1-x) \log P_X(x,q^2) \right]  \nonumber \\ 
&&+ \frac{ 2 M_B^2 \mathcal{H}^2}{9 \pi^2 f^2} \sum_X A_X \Bigg\{ \frac{1}{6} \log \frac{m_X^2}{\mu^2} +
\int_0^1 dx \, x(1-x) \Bigg[ 2 \log P_X(x,q^2)   \nonumber \\
&&\phantom{gfhjhgjhghg}- \frac{\Delta m_X P_X(x,q^2)}{\Delta^2 - m^2_X P^2_X(x,q^2)}  
\cR \left( \frac{\Delta}{m_X P_X(x,q^2)} \right)  \Bigg] \Bigg\}
.\end{eqnarray}

\section{\label{s:su2} $\D$ electromagnetic properties in $SU(2)$ flavor
         with non-degenerate quarks}
In this Section, we consider the case of $SU(2)$ 
flavor and calculate 
the electromagnetic moments and charge radii of the delta quartet. 
We keep the up and down valence quark masses non-degenerate and similarly 
for the sea-quarks. Thus the quark mass matrix reads
$m_Q^{SU(2)} = \diag(m_u, m_d, m_j, m_l, m_u, m_d)$.
Defining ghost and sea quark charges is constrained only by the 
restriction that QCD be recovered
in the limit of appropriately degenerate quark masses. 
Thus the most general form of the charge matrix is
\begin{equation}
  \cQ^{SU(2)} 
  = \diag\left(\frac{2}{3},-\frac{1}{3},q_j,q_l,q_j,q_l \right) 
.\end{equation}
The symmetry breaking pattern is assumed to be 
$SU(4|2)_L \otimes SU(4|2)_R \otimes U(1)_V
 \longrightarrow
 SU(4|2)_V \otimes U(1)_V$.
The baryon field assignments are analogous to the case of
$SU(3)$ flavor.
The nucleons are embedded as
\begin{equation} \label{eqn:SU2nucleons}
 \cB_{ijk}=\frac{1}{\sqrt{6}}\left(\e_{ij} N_k + \e_{ik} N_j\right)
,\end{equation}
where the indices $i,j$ and $k$ are restricted to $1$ or $2$ and 
the $SU(2)$ nucleon doublet is defined as
\begin{equation}
  N = \left(\begin{matrix} p \\ n \end{matrix} \right) 
\end{equation}
The decuplet field $\cT_{ijk}$, 
which is totally symmetric, 
is normalized to contain the $\Delta$-resonances 
$T_{ijk}=\cT_{ijk}$ with $i$, $j$, $k$ restricted to 1 or 2.
Our states are normalized so that $\cT_{111} = \D^{++}$.
The construction of the octet and decuplet baryons containing 
one sea or one ghost quark is analogous to the $SU(3)$ flavor
case~\cite{Beane:2002vq}
and we will not repeat it here.

The free Lagrangian for $\cB$ and $\cT$ is the one in 
Eq.~(\ref{eqn:L})
(with the parameters having different numerical values than
the $SU(3)$ case).  
The connection to QCD is 
detailed in~\cite{Beane:2002vq}.
Similarly, the Lagrangian describing the interaction of the 
$\cB$ and $\cT$ with the pseudo-Goldstone bosons is 
the one in 
Eq.~(\ref{eqn:Linteract}).  Matching it to the familiar one in QCD
(by restricting the $\cB_{ijk}$ and $\cT_{ijk}$ to the $qqq$ sector),
\begin{equation}
  \cL = g_{\D N} \left( \ol{T}{}^{kji}_{\nu} A_{il}^{\nu} N_j \e_{kl} 
                      + \text{h.c}
                 \right)
	+ 2 g_{\D \D} \ol{T}{}^\nu_{kji} S_\mu A^\mu_{il} T_{\nu,ljk} 
	+ 2 g_X  \ol{T}{}^\nu_{kji} S_\mu  T_{\nu,ijk} \tr (A^\mu)  
,\end{equation}
one finds at tree-level ${\mathcal C} = - g_{\D N}$ and $\mathcal{H} = g_{\D \D}$,
with $g_X = 0$. The leading tree-level operators which contribute to $\D$ electromagnetic
properties are the same as in Eqs.~\eqref{eqn:LDF}, \eqref{eqn:Lc}, and \eqref{eqn:Lnew}, of course the low-energy
constants have different values.

\begin{table}
\caption{The SU$(2)$ coefficients $A_X^T$ in \CPT\ and \PQCPT.}  
\begin{ruledtabular}
\begin{tabular}{l | c | c  c  c  c  c  c  c }

	   & \CPT\  &  \multicolumn{7}{c}{\PQCPT} \\
	                 & $\pi^{\pm}$   &   $uu$   &   $ud$   &  $dd$  &   $ju$   &   $lu$   & $jd$   & $ld$ \\
	 	
	\hline
	$\Delta^{++}$       & $1$  &  $-\frac{2}{3} + q_j$ & $\frac{1}{3} + q_l$ & $0$ & $\frac{2}{3} - q_j$ & $\frac{2}{3} - q_l$ & $0$ & $0$ \\
 
	$\Delta^{+}$        & $\frac{1}{3}$ & $-\frac{4}{9} + \frac{2}{3} q_j$ & $\frac{1}{3} q_j + \frac{2}{3} q_l$ & $\frac{1}{9} + \frac{1}{3} q_l$ & $\frac{4}{9} - \frac{2}{3} q_j$ & $\frac{4}{9} - \frac{2}{3} q_l$ & $-\frac{1}{9} - \frac{1}{3} q_j$ & $-\frac{1}{9} - \frac{1}{3} q_l$ \\

	$\Delta^{0}$        & $-\frac{1}{3}$ & $-\frac{2}{9} + \frac{1}{3} q_j$ & $-\frac{1}{3} + \frac{2}{3} q_j + \frac{1}{3} q_l$ & $\frac{2}{9} + \frac{2}{3} q_l$ & $\frac{2}{9} - \frac{1}{3} q_j$ & $\frac{2}{9} - \frac{1}{3} q_l$ & $-\frac{2}{9} - \frac{2}{3} q_j$ & $-\frac{2}{9} - \frac{2}{3} q_l$ \\

	$\Delta^{-}$        & $-1$ & $0$ & $-\frac{2}{3} + q_j$ & $\frac{1}{3} + q_l$ & $0$ & $0$ & $-\frac{1}{3} - q_j$ & $-\frac{1}{3} - q_l$ \\
\end{tabular}
\end{ruledtabular}
\label{t:su2}
\end{table} 

Evaluating the $\D$ electromagnetic properties at NLO in the chiral expansion yields expressions identical
in form to those above Eqs.~\eqref{eqn:F2}, \eqref{eqn:G1}, and \eqref{eqn:r_E} with the SU$(2)$ identifications made for $\mathcal{C}$
and $\mathcal{H}$ above. The SU$(2)$ coefficients $A_X^T$ appear in Table \ref{t:su2} for particular
$\D$--resonance states $T$. In the table, we have listed values corresponding to the loop meson that has mass $m_X$ 
for both \CPT\ and \PQCPT. Again, the \CPT\ coefficients can be used to find the $\D$-resonance charge radii in two-flavor QCD. 
These have not been previously calculated.

\end{document}